# Hardware-Efficient Template-Based Deep CNNs Accelerator Design

Azzam Alhussain
*College of Engineering and Computer Science*
*University of Central Florida*
Orlando, Florida, USA
mr.azzam@knights.ucf.edu

Mingjie Lin
*College of Engineering and Computer Science*
*University of Central Florida*
Orlando, Florida, USA
milin@ucf.edu

*Abstract*—Acceleration of Convolutional Neural Network (CNN) on edge devices has recently achieved a remarkable performance in image classification and object detection applications. This paper proposes an efficient and scalable CNN-based SoC-FPGA accelerator design that takes pre-trained weights with a 16-bit fixed-point quantization and target hardware specification to generate an optimized template capable of achieving higher performance versus resource utilization trade-off. The template analyzed the computational workload, data dependency, and external memory bandwidth and utilized loop tiling transformation along with dataflow modeling to convert convolutional and fully connected layers into vector multiplication between input and output feature maps, which resulted in a single compute unit on-chip. Furthermore, the accelerator was examined among AlexNet, VGG16, and LeNet networks and ran at 200-MHz with a peak performance of 230 GOP/s depending on ZYNQ boards and state-space exploration of different compute unit configurations during simulation and synthesis. Lastly, our proposed methodology was benchmarked against the previous development on Ultra96 for higher performance measurement.

## I. INTRODUCTION

Convolutional Neural Network (CNN) has achieved state-of-the-art accuracy in human-based image classification. For instance, FaceNet [1] has an astonishing accuracy of 99% for human face recognition in real-time applications. Besides, CNNs have different architecture designs, such as AlexNet [2], VGG16 [3], and ResNet [4], but mainly consist of convolutional, ReLU, pooling, flatten, Fully Connected (FC), and SoftMax layers as illustrated in Fig. 1. Krizhevsky *et al.* [2] has shown that 95% of the intensive computation workloads are comprised among convolutional and FC layers. Although the dimension and dataflow are considered the only difference between these layers, in which FC layers have a one-dimensional vector and convolutional layers usually have a 3-dimensional array. Nevertheless, advanced CNNs architecture becomes a challenging task and effort-hungry process to scale their workloads efficiently into one solution.

However, CNN's performance has increased dramatically the last couple of years at the cost of huge computational workload that is challenging for the Central Processing Units (CPUs). Therefore, Graphical Processing Units (GPUs) have been the most suitable integration unit to implement Deep Neural Networks (DNNs) on floating-points due to their highly optimized parallel processing cores, but at a significantly higher energy consumption which is not suitable for mobiles' and robots' vision applications. Alternatively, Field-Programmable Gate Array (FPGA) is an attractive acceleration platform which has a customizable internal structure that optimize the computation as GPUs and implements computer vision algorithms efficiently under fixed-points with lower energy consumption and minimum processing delay.

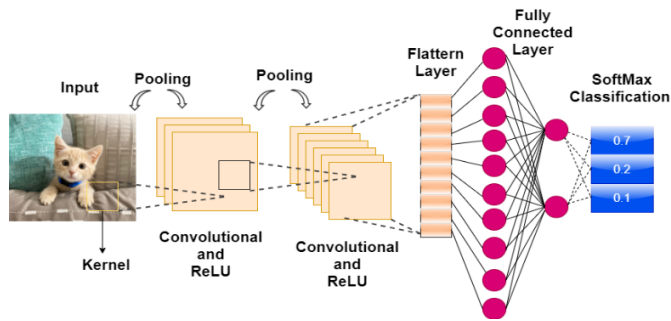

Figure 1. Traditional CNN architecture

The aim of this research is to find common patterns among two algorithms, create an HW/SW partitioning scheme, then develop an efficient and scalable accelerator on the Programmable Logic (PL) to compute the intensive operations of convolution and FC layers and gain higher performance. On the contrary, network initialization, pooling, normalization, SoftMax, and other layers are performed on the Processing System (PS). The contributions of this paper are as follows:

- Computed convolutional and FC layers operations in vector multiplication on a single on-chip compute unit.

- Utilized loop tiling transformation efficiently to construct the IP accelerator core.

- Demonstrated that the proposed methodology achieved superior performance up to 230 GOP/s under 200-MHz with minimum data execution time.

The rest of this paper is organized as follows: Section II provides background on Deep CNNs architecture and its hardware implementation. Section III discusses the methodology of the network architecture and the accelerator mechanism. Section IV describes the experimental tools and presents the results, and Section V concludes the paper.

## II. BACKGROUND

DNNs acceleration and deployment on FPGAs is an active research area compared with other edge computing platforms due to its efficiency and scalability. For the efficiency side, most researchers focused on optimizing the computation engine of the systolic array and neglected the external memory bandwidth, which resulted in lower performance. For this reason, Chakradhar *at el.* [5] discussed a separate bank of scratchpad DRAM to improve the computation. Authors in [6], [7] analyzed the data access patterns and used on-chip buffers to store the tiles of external data and maximize the reusability of BRAM. Additionally, *Guan et at.* [8] used data compression, while Hu *et al.* [9] proposed data quantization to overcome this obstacle.

Researchers in [10]-[14] reported major issues on the scalability side. First, different CNNs architecture have different layers' parameter that complicate the accelerator design. Second, some platforms are limited in scalability due to their restricted resources. However, the most recent work in line with our idea is proposed by Bjerge *et al.* [10], which implemented a 16-bit quantized CNN in 2.14 format of two bits integer and fourteen bits fractional using a PYNQ framework [15]. Nevertheless, this paper is qualitatively distinct from the past ones. Our proposed methodology addressed these issues and optimized performance, latency, and resource utilization, then benchmarked it against the previous developments.

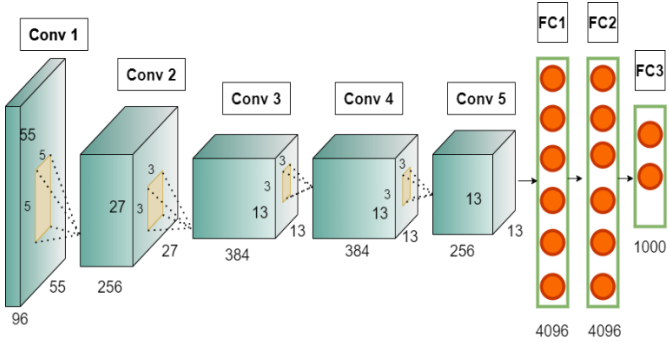

Figure 2. AlexNet network architecture

## III. METHODOLOGY

### A. Network Architecture

CNN architecture of AlexNet [2] shown in Fig. 2, VGG16 [3], and LeNet are used as a case study for the proposed template. All of these networks consist of convolutional and FC layers while the number and size of these layers vary across different operations. The algorithm of convolutional layers is illustrated as in the equation follows:

$$\forall\, row \in \{1,2,\dots,R\}$$
$$\forall col \in \{1,2,\dots,C\}$$
$$\forall co \in \{1,2,\dots,q\}$$
$$\forall ci \in \{1,2,\dots,p\}$$
$$\forall i \in \{3,5,\dots,\mathcal{K}\}$$
$$\forall j \in \{3,5,\dots,\mathcal{K}\}$$

$$OFM\,[row\!:\!col\!:\!co] = \sum\sum \begin{array}{l} IFM[s*row + i][s*col \\ + j][ci]\, W\,[co][ci][i][j]\end{array} \quad (1)$$

Rows ($R$) and columns ($C$) represent the image/matrices size, while the input channel ($p$) and output channel ($q$) are the third dimensions of input feature map ($IFM$) and output feature map ($OFM$) respectively and ($\mathcal{K}$) represents the kernel/filter size. The total number of convolutional operations are as in the equation follows:

$$No.\,of\,CONV\,Operations\ =\ \prod 2RC\,p\,q\,\mathcal{K}^2 \quad (2)$$

On the other hand, the algorithm of FC layers is illustrated as in the equation follows:

$$\forall co \in \{1,2,\dots,q\,\}$$
$$\forall ci \in \{1,2,\dots,p\}$$

$$OFM\,[co] = \sum\sum IFM[ci] \times W[co][ci] \quad (3)$$

Where the operations of FC Layers have one dimensional data on the input and output neurons which has less processing tasks compared to the convolutional layers. The total number of FC operations is as in the equation follows:

$$No.\,of\,FC\,Operations = \prod 2pq \quad (4)$$

And as previously mentioned, these layers are operationally expensive in terms of computation and latency. For this reason, it is pertinent to map them into the PL part of ZYNQ. The previous equations represent symmetrical operation while the dynamics and dataflow of the layers are different. The proposed methodology finds a common dataflow pattern for an optimized accelerator design.

### B. Loop Tiling Transformation

Equation. 1 represents layer values of $R$, $C$, $p$, $q$, and $\mathcal{K}$ as variable values in which using these values to perform direct implantation leads to an inefficient accelerator design. Hence, loop tiling is performed, which converts loops into fixed points/blocks. It is also represented in the same algorithm and uses the tile size as ($\mathcal{T}$) for $R$, ($\mathbb{C}$) for $C$, ($\mu$) for $p$, and ($\tau$) for $q$. Thus, we can transfer a fixed amount of data from the external memory (DRAM) to on-chip memory (BRAM). Once the data is cashed into BRAM, fixed computations are performed by the accelerator. On the contrary, loop tiling for the FC layers uses the tile size as ($\lambda$) for $p$ and ($\Omega$) for $q$.

In the accelerator design, the FC layers have larger vector values in the input and output channels compared to the convolutional layers. Hence, different sizes of tile are chosen. However, choosing the same size of tiles for the input and output channels resulted in performance reduction. Lastly, the overall selection of these tiles can result in maximum resource utilization and lower latency.

### C. Accelerator Design

The proposed template-based vector design is illustrated in Fig. 3, shown the loop tiling factors of $\mathcal{T}$, $\mathbb{C}$, $\mu$, $\tau$, $\lambda$, and $\Omega$ that determine the on-chip buffers. Due to the size difference of input and output channels in convolutional and FC layers, different sizes of tiles are used and prompted to use dedicated buffers for both types of layers. This method resulted in more resource utilization and overcome the reading and writing overhead latency owing to the multi-dimensional array. Additionally, those dedicated buffers are being used in the size of weights in both layers which allow the design to have better efficiency and lower latency at the cost of more resource utilization.

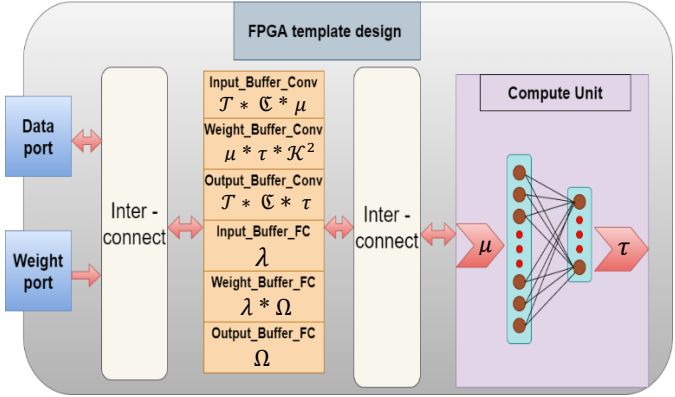

Figure 3. The proposed FPGA template-based design

First, data is cashed on-chip in these buffers using two data ports where one can be used for read/write of ($IFM$) and ($OFM$), and the other one can be used only for reading the weights. Those ports are memory-mapped (M-AXI), which enables a burst transfers and improv the external memory bandwidth. The input and weight buffer are partitioned in dimension $\tau$ to improve the design latency as multiple reads and writes are possible on arrays. This proposed compute unit is designed to do a dot-product between $\mu$ and $\tau$ input/output neurons respectively.

The template design has a scheduling mechanism among the interconnect, which orchestrate control logic and dataflow. Furthermore, the ping pong data transfer method is used on the input, weight, and output buffers to ensure a simultaneous data transfer happens from DRAM to on-chip buffers and then from on-chip buffers to the compute unit. Lastly, equations of the total operations performed by IP accelerator (3) and performance measurement in Giga Operations per second (GOP/s) (4) are as follows:

$$Accelerator\ IP = \mathcal{I}_\mathcal{P} = \prod 2\ \mathcal{T}\ \mathfrak{C}\ \mu\ \tau\ \mathcal{K}^2 \qquad (3)$$

$$Performance\ GOP/s = \mathcal{G}_{\mathcal{PS}} = \frac{\mathcal{I}_\mathcal{P}}{Latency} \qquad (4)$$

### D. Dataflow Modeling

Dataflow modeling is done concerning the architectural details of on-chip buffers and the compute unit. Convolutional algorithm in equation. 1 is based on window operation in which $\mathcal{K} * \mathcal{K}$ weight window is convolved with $\mathcal{K} * \mathcal{K}$ patch of the input pixel of the ($IFM$). The sum of these operations is resulted in the ($OFM$) at a particular index. This straightforward approach has a complex data pattern on the FPGA which produce poor architecture design. On the contrary, the FPGA can parallelize the workload, so layers dataflow is simplified and present low dependency from the on-chip buffers to the compute unit.

The dataflow modeling of convolutional layers in Fig. 4 is shown $\mathcal{T} * \mathfrak{C} * \mu$ as an input neuron, $\mathcal{T} * \mathfrak{C} * \tau$ as an output neuron, and $\mu * \tau * \mathcal{K}^2$ as a weight value where all of them are cashed into input, output, and weight buffers for on-chip processing. First, the data is moved from the input and weight buffers to the compute unit to perform dot product which and resulted in a written values in the output buffer. Then, the dataflow occurs in a form of vector values across those channels. All ($IFM$) values across channels (0, 1, 2,...,$\mu$-1) are read starting from the index (0,0) to the last index ($\mathcal{T}$-1, $\mathfrak{C}$-1) and then transferred alongside with the weights values to the compute unit. After that, the compute unit performs dot product along the channel dimension of ($OFM$) and resulted in the output vector (0, 1, 2,...,$\tau$-1). This process is continuously repeated for a spatial location of $\mathcal{K} * \mathcal{K}$ on ($IFM$) and then stored on ($OFM$) to achieve high parallelism in the dimension of input and output channels ($\mu$, $\tau$) and reduce data dependency for reading and writing among those buffers.

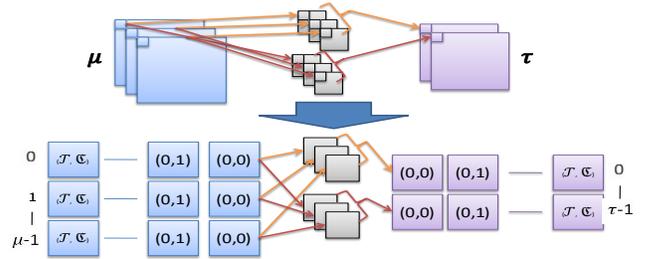

Figure 4. Depiction of Convolution dataflow and computation on FPGA

On the other hand, FC layers working principle is illustrated in Fig. 5 shown $\lambda$ input neurons and $\lambda * \Omega$ weight values are cashed on the BRAM buffers. These values are too large to be processed by the compute unit at once. As a result, another set of loop tiling/block is introduced for the FC layers which break ($\lambda$, $\Omega$) data into smaller ($\mu$, $\tau$) sizes. The input size $\mu$ and the weights size $\mu * \tau$ are transferred to the compute unit while the output size $\tau$ is written back to the output buffer. This method ensure that the entire input vector is processed efficiently on the same compute unit.

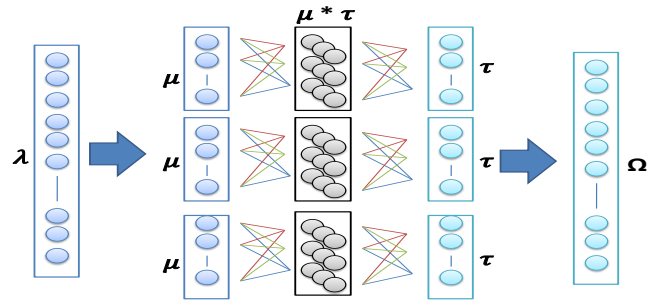

Figure 5. Depiction of Fully Connected dataflow and computation on FPGA

### E. Scalability and Efficiency

The proposed accelerator utilized a pre-trained model with 16-bit fixed-point quantization in 2.14 format. Nevertheless, the tiling size of convolution and FC layers determines the optimum performance of the template. We used a trial-based method to fine-tune the accelerator for higher efficiency and better scalability across various SoC-FPGAs ZYNQ boards and CNNs architecture. The parameters were randomly selected, and the design was simulated until the resources and latency were met. After sets of trial and error, we found that the tile factor of $\mu*\tau$ affect the performance directly in which this proposed template achieved higher performance when $\tau$ is approximately twice $\mu$ under resource constraints. As a result, the accelerator can run advanced CNN architecture such as ResNet-50, SSD, MobileNets and YOLO of any version. Since all of these networks have same type of layers, the proposed methodology can map any CNN architecture.

## IV. EXPERIMENTAL SETUP AND RESULTS

### A. Development Environments

The accelerator design was simulated and synthesized using Vivado High-Level Synthesis (HLS) (2019.2), and utilized pre-trained models from PyTorch Model Zoo [16]. It was tested with AlexNet [2], VGG-16 [3], and LeNet architectures and can work with any advanced CNNs network.

### B. Results

The proposed template can operate under 200MHz, and achieve superior performance of up to 230 GOP/s. Table 1 reported the resource utilization and performance measurement of AlexNet network demonstrated on Ultra96, ZCU104, and ZCU102. The BRAM and DSP are directly dependent on the tile size of $\mathcal{T}$ and $\mathfrak{C}$, and the number of dot-products in the compute unit, while FF and LUT are used to control the logic gates and state machine for running the loops and controlling the dataflow. Finally, our accelerator was benchmarked against the previous development [10] on Ultra96 and achieved higher performance and lower latency, as reported in Table 2.

Table 1. Resource utilization and performance measurement

| Device | Ultra96 | ZCU104 | ZCU102 |
|---|---|---|---|
| Compute Unit $\mu*\tau$ | 12 x 24 | 20 x 30 | 20 x 55 |
| Flip-Flops | 23.5k (16%) | 46k (10%) | 139k (25%) |
| LUTs | 15.6k (22%) | 24k (10%) | 57k (20%) |
| BRAM | 332 (76%) | 594 (95%) | 1.7K (95%) |
| DSP Slices | 334 (92%) | 586 (33%) | 1.7K (67%) |
| Performance | 51 GOP/s | 107 GOP/s | 230 GOP/s |
| Frequency | 169 MHz | 198 MHz | 167 MHz |

Table 2. Benchmark and comparision

| Device | Ultra96 | |
|---|---|---|
| Development | Previous method [10] | Proposed method |
| Max frequency | 170 MHz | 169 MHz |
| Bit width | 16 | 16 |
| Performance | 31 GOP/s | 51 GOP/s |
| Latency (ms) | 4.6 | 0.174 |
| Power (w) | 3.55 | 4.7 |

## V. CONCLUSION AND FUTURE WORK

The proposed template efficiently utilized the loop tiling and dataflow modeling for optimized accelerator design. As a result, a range of 1.3x - 1.7x higher performance was achieved along with a minimal layer of execution time when compared to the previous development [10]. The analysis and simulation results proved to be optimistic and can be extended to create a complete framework. This will allow the community to use our open-source project and search an efficient implementation for real-time applications [17].